\documentclass[fleqn,usenatbib]{mnras}

\usepackage{newtxtext,newtxmath}

\usepackage[T1]{fontenc}

\DeclareRobustCommand{\VAN}[3]{#2}
\let\VANthebibliography\thebibliography
\def\thebibliography{\DeclareRobustCommand{\VAN}[3]{##3}\VANthebibliography}

\usepackage{graphicx}	
\usepackage{amsmath}	

\title[XTE J1701-462]{Radio observations of the 2022 outburst of the transitional Z-Atoll source XTE J1701-462}

\author[K. V. S. Gasealahwe et al.]{K. V. S. Gasealahwe,$^{1,2}$\thanks{E-mail: kelebogile@saao.ac.za}
I. M. Monageng,$^{1,2}$
R. P. Fender,$^{1,3}$
P. A. Woudt,$^{1}$
A. K. Hughes,$^{4}$
S. E. Motta,$^{3,5}$
\newauthor J. van den Eijnden,$^{6}$
P. Saikia,$^{7}$
E. Tremou,$^{8}$
\\
$^{1}$Department of Astronomy, University of Cape Town, Private Bag X3, 7701 Rondebosch, South Africa\\
$^{2}$South African Astronomical Observatory, P.O. Box 9, 7935
Observatory, South Africa \\
$^{3}$Department of Physics, University of Oxford, Denys Wilkinson Building, Keble Road, Oxford OX1 3RH, UK \\
$^{4}$Department of Physics, University of Alberta, CCIS 4-181, Edmonton, AB T6G 2E1, Canada \\
$^{5}$Istituto Nazionale di Astrofisica, Osservatorio Astronomico di Brera, via E. Bianchi 46, 23807 Merate (LC), Italy \\
$^{6}$Department of Physics, University of Warwick, Coventry CV4 7AL, UK \\ 
$^{7}$Center for Astrophysics and Space Science (CASS), New York University Abu Dhabi, PO Box 129188, Abu Dhabi, UAE \\
$^{8}$National Radio Astronomy Observatory, Socorro, NM 87801, USA }
\date{Accepted XXX. Received YYY; in original form ZZZ}

\pubyear{2015}

\begin{document}
\label{firstpage}
\pagerange{\pageref{firstpage}--\pageref{lastpage}}
\maketitle

\begin{abstract}
XTE J1701-462 is a neutron star low mass X-ray binary (NS LMXB) discovered in 2006 as the first system to demonstrate unambiguously that the `Atoll' and `Z' classes of accreting neutron stars are separated by accretion rate. 
Radio observations during the 2006/7 outburst provided evidence for the formation of a relativistic jet, as now expected for all accreting neutron star and black hole X-ray binaries at high accretion rates. The source entered a new outburst in 2022, and we report 29 observations made with the MeerKAT radio telescope. The first radio detection was on the 16th September 2022, we continued detecting the source until mid-December 2022. Thereafter, establishing radio upper limits till 25 March 2023.  We present the radio analysis alongside analysis of contemporaneous X-ray observations from MAXI. The radio light curve shows evidence for at least three flare-like events over the first hundred days, the most luminous of which has an associated minimum energy of $1\times10^{38}$\,erg. We provide a detailed comparison with the 2006/7 outburst, and demonstrate that we detected radio emission from the source for considerably longer in the more recent outburst, although this is probably a function of sampling.  We further constrain the radio emission from the source to have a polarisation of less than 9\% at the time of 2022 IXPE detection of X-ray polarisation. Finally, we place the source in the radio -- X-ray plane, demonstrating that when detected in radio it sits in a comparable region of parameter space to the other Z-sources.
\end{abstract}




\section{Introduction}
Low-mass X-ray binaries (LMXBs) consist of a compact object that could be a black hole (BH) or a neutron star (NS) and low-mass star as a binary companion. In these systems the compact object may accrete matter from the companion via an accretion disc showing X-ray outbursts as a result of the mass transfer between the objects. During such outbursts the sources can be detected in the radio due to the synchrotron emission from outflowing jets. 
The collimated compact jets follow a non-linear radio -- X-ray luminosity correlation, where $L_{\rm R} \propto L^{\beta}_{\rm X}$ with $\beta \approx$ 0.5 -- 0.7 \citep{2003A&A...400.1007C,2013MNRAS.428.2500C,2018MNRAS.478L.132G}, which extends to supermassive BHs through the fundamental plane of BH activity \citep{2003MNRAS.345.1057M,2004A&A...414..895F,2015MNRAS.450.2317S,2018MNRAS.477.2119S}, suggesting scale invariance of compact jets.
The radio -- X-ray correlation for BH and NSs is expected to differ considerably due to the physical nature of the compact objects. This is largely due to radiative inefficient accretion assumed for BHs due to gravitational potential being lost across the event horizon. Since matter accretes onto the NS surface it is expected to be more radiatively efficient such that the radio -- X-ray correlation is steeper. Studies over the past two decades suggest however, that the radiative efficiency may be similar between the two classes of objects with sources following parallel tracks on the radio:X-ray correlation plane \citep{2015ApJ...809...13D,2020MNRAS.492.2858G}. 
\\\\
Z- and Atoll sources are low magnetic field NS LMXBs \citep{1989A&A...225...79H,2002ApJ...568L..35M}. The systems are classified as Z- or Atoll based on the patterns traced out by their X-ray colour-colour diagrams (CCDs) or their hardness intensity diagrams (HID) (see examples of CCD and HID for Z- and Atoll sources in Figures from \citealt{1989ESASP.296..203V,1989A&A...225...79H,2003ApJ...596.1155V,2007ApJ...656..420H,2009ApJ...696.1257L}). The luminosities reached by the Z-sources are close to and can exceed the Eddington luminosities ($L_{\rm{EDD}}$) and they trace out roughly Z-shaped tracks in CCDs/HIDs within hours to days \citep{2009ApJ...696.1257L} during the outbursts. The Z track is divided into the an upper, diagonal, and lower branch, which are called horizontal, normal, and flaring branches (HB/NB/FB), respectively. 
Furthermore, \cite{1989A&A...225...79H} groups Z-sources into Sco- (eg. GX 17+2, GX 349+2 and Sco X-1) or Cyg- (eg. Cyg X-2, GX 340+0, GX 5-1) like sources. In the former, the HB is nearly vertical or incomplete and the FB has strong flaring with high count rates, while in the latter the HB is almost horizontal in the HID and count rates decrease on the FB \citep{2007ApJ...656..420H,2014MNRAS.443.3270M}. The Atoll sources span a lower and larger luminosity range (0.001–0.5 $L_{EDD}$ \citealt{2007ApJ...656..420H}), tracing out their patterns in CCDs/HIDs on longer timescales (from days to weeks) \citep{2009ApJ...696.1257L}. The X-ray spectra of Atoll sources are soft at high luminosities, but hard when they are faint. The HID patterns of the Atolls are called the extreme island, island, and banana states and \cite{2009ApJ...696.1257L} refers to the Atoll patterns as hard, transitional, and soft states, respectively. 
Comparing Z- and Atoll sources on the radio -- X-ray plane will provide a better understanding of the relationship or transitional nature between Z and Atoll sources.
\\\\
A source can show transitions between Z and Atoll source signatures. XTE J1701-462 (hereafter XTE J1701) is one such example, while Cir X-1 has also shown these traits \citep{1995A&A...297..141O,1998ApJ...506..374S}. 
XTE J1701 is a transient source which showed characteristics of Z-source behaviour when it was first discovered in 2006 \citep{2007ApJ...656..420H}. It is the first NS transient to show all the characteristics of a Z-source and has a distance estimate of 8.8 $\pm$ 1.32\,kpc \citep{2009ApJ...696.1257L}. In the first 10 weeks of its 600 day discovery outburst, XTE J1701 transformed from a Cyg- like source into a Sco- like Z-source, and during the decay it evolved further into an Atoll source. The 2006/7 outburst of XTE J1701 spans a large range in luminosity, from super-Eddington towards quiescence, suggested to imply significant changes in the mass accretion rate as it transitions from Z- to Atoll \citep{2011AJ....142...34D}. The lightcurve of the XTE J1701 2006/7 outburst was broken down into stages (I -- IV) \citep{2009ApJ...696.1257L} . The last three stages show behaviour of Sco-like sources, and in the fourth stage (IV) it is observed that the HB and NB are no longer present. The CCDs/HIDs resemble those of the bright persistent GX Atoll sources (GX 9+1, GX 9+9, and GX 3+1) which \cite{2009ApJ...696.1257L} classified into the NB/FB vertex and the FB for XTE J1701. \cite{2007MNRAS.380L..25F} discussed a large scale structure seen in the radio which may be associated with the source; however, \cite{2023MNRAS.521.2806G} reported that the structure is likely an extragalactic background source. 
\\\\
The source has since gone into a new outburst during 2022/23 which lasted over 200 days; reports from that outburst include IXPE observations \citep{2023A&A...674L..10C} which were taken on 29-30 September and 8-9 October 2022. \cite{2023A&A...674L..10C} reported $\sim$ 4.6 $\pm$ 0.4\% linear polarisation of the X-rays from the source during the upper and intermediate horizontal branch, the highest value of polarisation found for this class of source. 
\\\\
In this paper we discuss the new (2022/23) outburst in the radio band with the MeerKAT, comparing it to the first outburst to develop a better understanding of the transitional nature between Z and Atoll. We present a description of the observations in Section 2 and report the results in Section 3. In Section 4.1 we compare the 2006/7 and 2022/23 outbursts, the polarisation constraint in Section 4.2, the in-band spectral index and minimum energy of the flare seen in the 2022/23 lightcurve are reported in Section 4.3. Furthermore, we discuss the position of XTE J1701 on the radio -- X-ray correlation in Section 4.4.

\begin{table*}
\caption{The table below lists the MeerKAT radio flux densities at 1.28\,GHz and assuming a flat spectral index the luminosities at 5\,GHz for the 2022/23 outburst. Similarly we list the X-ray flux and luminosities from MAXI calculated with N$_H$ 2.59 and photon index $\Gamma = 1.82$. The MeerKAT radio spectral indices determined for the detections are are also shown in the table below.}
\begin{tabular}{l l l l l l}\hline
      MJD	&	F$_{\rm{R}}$ [mJy/beam]	&	L$_{\rm{R}}\times 10^{29}$ [erg/s]	&	F$_{\rm{X}}\times 10^{-9}[\rm{ergs/cm^2/s}]$	&		L$_{\rm{X}}\times 10^{37}$ [erg/s] & Spectral index ($\alpha; S_{\nu}\propto \nu^{\alpha}$)	\\	\hline
59829.50	& &	&	6.42	$\pm$	0.27	&	5.95	$\pm$ 0.25	\\		
59829.66	&	<0.07				&	<0.34		\\		
59831.50	& &	&	8.73	$\pm$	0.36	&	8.09	$\pm$ 0.33	\\		
59831.71	&	<0.07				&	<0.32		\\		
59838.72	&	0.34	$\pm$	0.02		&	1.57	$\pm$	0.09 & & & 0.26 $\pm$ 0.46\\		
59839.50	& &	&	10.50	$\pm$	2.05	&	9.73	$\pm$ 1.90	\\		
59842.50	& &	&	15.70	$\pm$	0.81	&	14.60	$\pm$ 0.75	\\		
59842.60	&	2.62	$\pm$	0.03		&	12.14	$\pm$	0.14 & & & -0.21 $\pm$ 0.26\\		
59846.50	& &	&	16.40	$\pm$	1.06	&	15.20	$\pm$ 0.98	\\		
59846.74	&	3.98	$\pm$	0.02		&	18.44	$\pm$	0.10 & & & -2.53 $\pm$ 0.20\\		
59852.54	&	1.71	$\pm$	0.03		&	7.90	$\pm$	0.13 & & & -0.08 $\pm$ 0.59\\		
59859.50	& &	&	21	$\pm$	0.60	&	19.50	$\pm$ 0.56	\\		
59859.70	&	1.31	$\pm$	0.02		&	6.10	$\pm$	0.11 & & & -0.84 $\pm$ 0.25 \\		
59867.48	&	0.92	$\pm$	0.02		&	4.26	$\pm$	0.10 & & & -0.36 $\pm$ 0.21 \\		
59867.50	& &	&	19.60	$\pm$	0.54	&	18.20	$\pm$ 0.50	\\		
59873.50	& &	&	13.30	$\pm$	0.29	&	12.30	$\pm$ 0. 26	\\		
59873.61	&	1.32	$\pm$	0.024		&	6.12	$\pm$	0.11 & & & -0.16 $\pm$ 0.22\\		
59880.50	& &	&	11.30	$\pm$	0.37	&	10.50	$\pm$ 0.34	\\		
59880.64	&	<0.07				&	<0.34		\\		
59889.50	& &	&	7.75	$\pm$	0.39	&	7.18	$\pm$ 0.37	\\		
59889.68	&	2.53	$\pm$	0.03		&	11.72	$\pm$	0.13 & & & -1.07 $\pm$ 0.21\\		
59894.64	&	<0.07				&	<0.32		\\		
59895.50	& &	&	6.31	$\pm$	0.51	&	5.85	$\pm$ 0.47	\\		
59901.50	& &	&	12.70	$\pm$	0.74	&	11.80	$\pm$ 0.69	\\		
59901.58	&	0.80	$\pm$	0.02		&	3.71	$\pm$	0.11 & & & -0.34 $\pm$ 0.19\\		
59909.50	& &	&	15.30	$\pm$	0.89	&	14.20	$\pm$ 0.82	\\		
59909.58	&	2.23	$\pm$	0.03		&	10.33	$\pm$	0.13  & & & -0.74 $\pm$ 0.33\\		
59915.50	& &	&	12.80	$\pm$	0.69	&	11.90	$\pm$ 0.64	\\		
59915.62	&	0.35	$\pm$	0.02		&	1.62	$\pm$	0.11 & & & -0.12 $\pm$ 0.80\\		
59921.50	& &	&	7.17	$\pm$	0.94	&	6.64	$\pm$ 0.87	\\		
59922.58	&	<0.08				&	<0.35		\\		
59932.35	&	<0.07				&	<0.34		\\		
59934.50	& &	&	9.86	$\pm$	0.44	&	9.14	$\pm$ 0.41	\\		
59939.31	&	0.22	$\pm$	0.03		&	1.02	$\pm$	0.12 \\		
59939.50	& &	&	11.40	$\pm$	0.38	&	10.6	$\pm$ 0.35	\\		
59946.44	&	<0.08				&	<0.36		\\		
59946.50	& &	&	10.50	$\pm$	0.48	&	9.73	$\pm$ 0.45	\\		
59951.38	&	<0.09				&	<0.36		\\		
59951.50	& &	&	9.40	$\pm$	0.26	&	8.71	$\pm$ 0.24	\\		
59958.38	&	<0.08				&	<0.35		\\		
59958.50	& &	&	6.51	$\pm$	0.38	&	6.03	$\pm$ 0.35	\\		
59965.42	&	<0.07				&	<0.33		\\		
59965.50	& &	&	6.10	$\pm$	0.68	&	5.65	$\pm$ 0.63	\\		
59972.34	&	<0.14				&	<0.67		\\		
59972.50	& &	&	7.13	$\pm$	0.27	&	6.61	$\pm$ 0.25	\\		
59979.25	&	<0.14				&	<0.66		\\		
59979.50	& &	&	9.16	$\pm$	0.58	&	8.49	$\pm$ 0.54	\\		
59986.38	&	<0.08				&	<0.36		\\		
59986.50	& &	&	6.20	$\pm$	0.53	&	5.75	$\pm$ 0.49	\\		
59994.50	& &	&	5.90	$\pm$	2.82	&	5.47	$\pm$ 2.61	\\		
59996.23	&	<0.14				&	<0.65		\\		
60001.22	&	<0.14				&	<0.65		\\		
60004.50	& &	&	4.28	$\pm$	0.31	&	3.97	$\pm$ 0.29	\\		
60015.25	&	<0.07				&	<0.33		\\		
60015.50	& &	&	1.80	$\pm$	0.17	&	1.67	$\pm$ 0.16	\\		
60028.04	&	<0.07				&	<0.33		\\		
60028.50	& &	&	0.01	$\pm$	0.10	&	0.01	$\pm$ 0.10	\\\hline											
\end{tabular}
\label{tab:table}
\end{table*}
\section{Observations}
\subsection{MeerKAT observations}
The latest outburst of XTE J1701 triggered follow-up observations with MeerKAT at a central frequency of 1.28\,GHz and bandwith of 856\,MHz. This was done through the ThunderKAT Large Survey Programme \citep{2016mks..confE..13F}. XTE J1701 was observed for 29 epochs nearly weekly over a span of seven months from 2022 to 2023. The observations started on the 7 September 2022 (MJD 59829.50) using an average of 61 dishes for a duration of 15 minutes each observation. The semi-automated pipeline, {\textsc{oxkat}}\footnote{for more details see, \url{https://github.com/IanHeywood/oxkat}} \citep{2020ascl.soft09003H} was used to reduce the data with observation calibrators J1939-6342 (bandpass, flux calibrator) and J1744-5144 (phase, amplitude calibrator). The first and second generation process of {\textsc{oxkat}} were used for this data, using {\textsc{casa}} \citep{2007ASPC..376..127M} to perform the averaging, flagging steps and do cross- and self-calibration. After the first generation process was complete, the visibility and gain solution plots were inspected and deemed satisfactory, so that the data were then flagged and imaged using the {\textsc{tricolour}} and {\textsc{wsclean}} packages respectively \citep{2014MNRAS.444..606O}. Thereafter, the second generation processing was initiated to do self calibration, plot gain solutions and a second output of images were produced. The flux densities were then determined using {\textsc{pybdsf}} \citep{2015ascl.soft02007M}.
\\\\
Additionally, we were interested in constraining the polarisation properties of the source during a brief period of $\sim$\,mJy-level flaring between 30 September 2022 and 7 October 2022. We corrected for on-axis instrumental polarisation (which converts unpolarised flux into polarised flux) using the unpolarised bandpass calibrator J1939-6342\footnote{Polarisation calibration was done with a modified version of \textsc{oxkat}. The calibration routine is available here: \url{https://github.com/AKHughes1994/polkat}}. Our observations did not include a strongly polarised calibrator, preventing us from correcting the instrumental cross-hand phase. The uncorrected cross-hand phase can cause a circular-to-linear polarisation conversion but does not modify the total amount of polarised flux \citep[see equations 16--19 in, ][]{2017AJ....154...54H}. Therefore, while we cannot measure the relative contributions from circular and linearly polarized emission, we can measure the total polarised flux. 
\\\\
Post-calibration, we produced full $I$, $Q$, $U$, and $V$ Stokes images using {\textsc{wsclean}}. The Stokes parameters correspond to four fluxes that describe the polarisation state, where $I$ is the total flux, $Q$ and $U$ are related to the linear polarisation, and $V$ is the circularly polarized flux; the total polarised flux is $P = \sqrt{Q^2 + U^2 + V^2}$, and the polarization fraction is $P/I$. We do not detect a polarised component spatially coincident with XTE J1701. To measure upper limits, we used the {\textsc{casa}} task \texttt{imfit} to extract the polarised flux by fitting a Gaussian component fixed at the position of the source (i.e., we performed forced aperture photometry). Given that the source is unresolved, we also fixed the shape of the Gaussian to that of the synthesized beam. Using our forced aperture results, we calculated 99.7$\%$ ($3\sigma$) upper limits on the polarised flux (and polarisation fraction) following the prescription from \citet{2006PASP..118.1340V}. We present our polarisation upper limits in Section 4.2. 

\begin{figure*}
    \centering
    \includegraphics[scale=0.45]{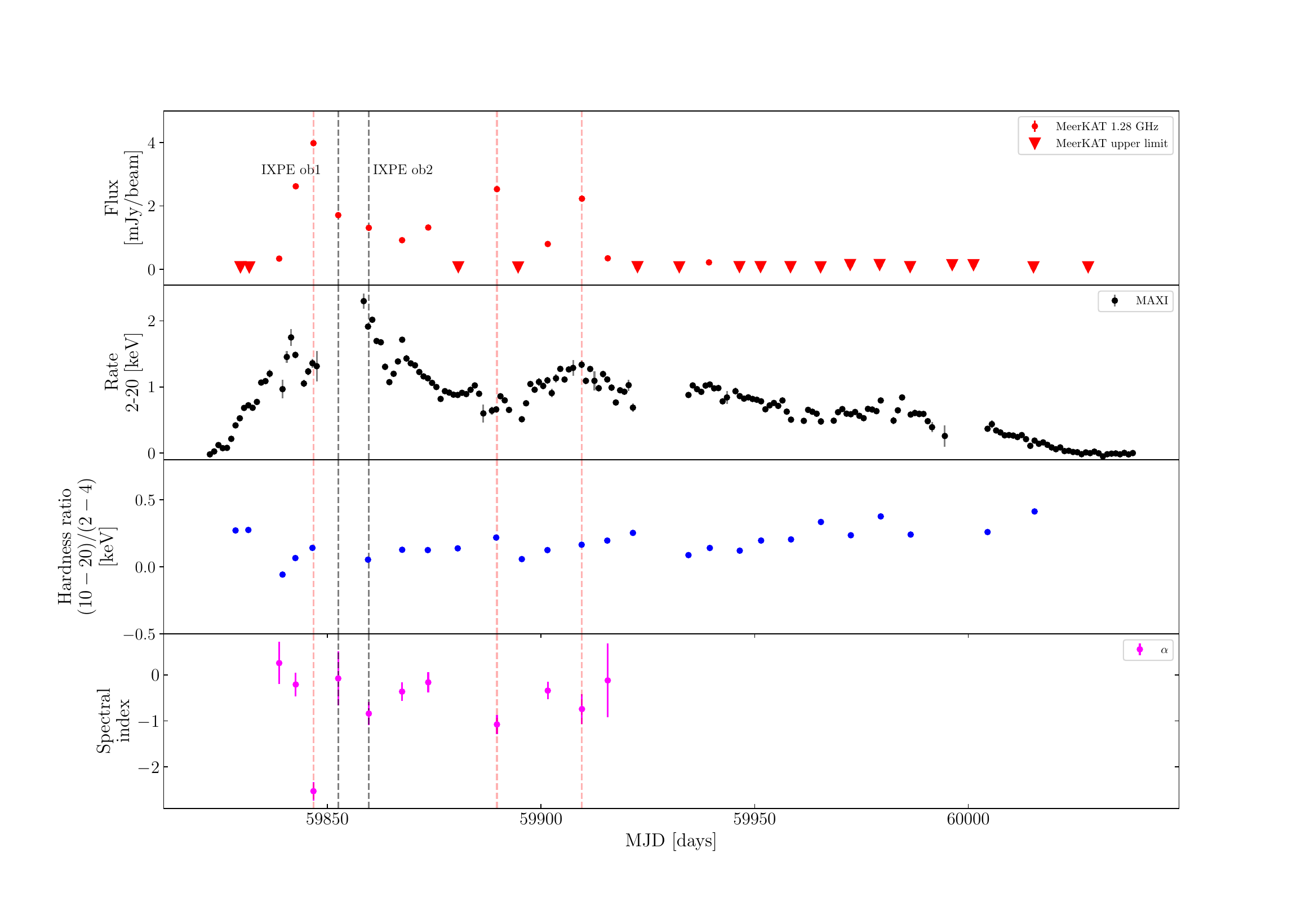}
    \caption{The radio (MeerKAT), X-ray (MAXI), hardness intensity (10-20)/(2-4) keV and radio (MeerKAT) spectral index lightcurves of the 2022/23 outburst of XTE J1701 in the first, second, third and bottom panel respectively. The IXPE observations taken from \protect\cite{2023A&A...674L..10C} are indicated with the grey dashed lines and are shown to coincide with our radio detections. The red dashed lines run through the peaks of the radio flares and the radio spectral indices are shown for all the radio detections except the final one from MJD 59939.31 since the source was not detected in several of the lower band frequencies.  }
    \label{fig:J1701_2022lightcurve}
\end{figure*}

\subsection{MAXI observations}
In order to contextualize the radio monitoring with X-ray observations of XTE J1701, we utilized data from the \textit{Monitor of All-Sky X-ray Image} \citep[\textit{MAXI};][]{matsuoka2009} Gas Slit Camera (GSC). We extracted the daily monitoring light curve of the target in the 2-20 keV band from the public \textit{MAXI} website\footnote{\url{https://maxi.riken.jp/top/index.html}}. For days with a MeerKAT radio observation, we also calculate the X-ray hardness and estimate the X-ray flux. The hardness is calculated by dividing the X-ray count rates in the two sub-bands between 10-20 keV and 2-4 keV. The flux, on the other hand, is estimated by converting the 4-10 keV count rate to the unabsorbed 1-10 keV X-ray flux using WebPIMMS\footnote{\url{https://heasarc.gsfc.nasa.gov/cgi-bin/Tools/w3pimms/w3pimms.pl}}. For this conversion, we assumed an absorbed power law spectral shape with index $\Gamma = 1.82$ and absorption column $N_H = 2.59\times10^{22}$ cm$^{-2}$, as derived in this outburst from observations with the \textit{Neil Gehrel's Swift Observatory} \citep{chandra2022}. 

\section{Results} 
 During the 2022/23 outburst MeerKAT observed XTE J1701 nearly weekly for 29 epochs. We measured 3$\sigma$ upper limits for the first two observations (see Table~\ref{tab:table}) and 0.34 $\pm$ 0.02\,mJy/beam for the first detection. The lightcurve (Figure~\ref{fig:J1701_2022lightcurve}) traces the outburst in the radio and the X-ray in the first and second panel respectively. The hardness ratio (10 - 20)/(2 - 4)\,keV  is shown in the third panel, where the points only represent the instances of quasi-simultaneity ($\Delta$ t = $\pm$ 1\,day ) between the MAXI and MeerKAT observations. In the fourth panel of Figure~\ref{fig:J1701_2022lightcurve} we display the in-band 
 radio spectral index over the period of the outburst, where the points are defined by the radio detections. The spectral indices are the slopes determined from 8 frequency bands; 0.91, 1.02, 1.12, 1.23, 1.34, 1.44, 1.55, 1.66\,GHz (see Figure~\ref{fig:inbandspec}) and at MJDs 59838.72 and 59915.61 the source was not detected in the lower bands but with at least three points we are able to determine a slope for the spectral index. Throughout the outburst we observe the source to go into multiple reflares in the radio and the X-ray. The radio flares peak at 3.98 $\pm$ 0.022\,mJy/beam (MJD = 59846.74), 2.53 $\pm$ 0.029\,mJy/beam (MJD = 59889.68) and 2.23 $\pm$ 0.027\,mJy/beam (MJD =  59909.58). After the first three radio flares MeerKAT did not detect the source after 4 months of observations but XTE J1701 continued detection in the X-ray (see second panel in Figure~\ref{fig:J1701_2022lightcurve}). 


\begin{figure*}
    \centering
    \includegraphics[scale=0.43]{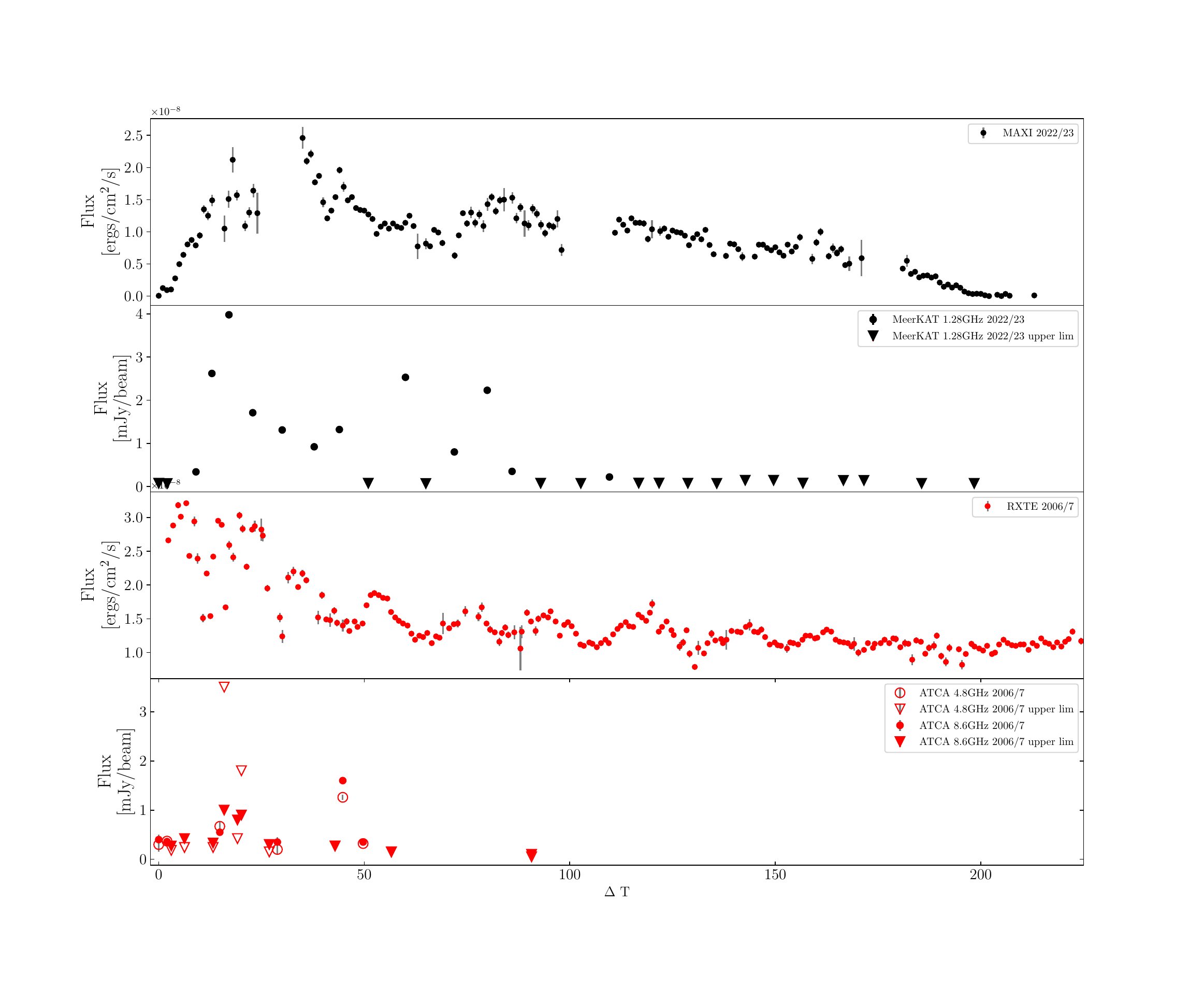}
 \caption{We compare the radio and X-ray lightcurves of XTE J1701 for the 2022/23 (black lightcurves) and 2006/7 (red lightcurves) outbursts. T=0 is 59829.65 and 53757.91 for the 2022/23 and 2006/7 outbursts, respectively. In the second and fourth panels for the radio lightcurves, the 3$\sigma$ upper limits are represented by the upside down triangles. }
    \label{fig:J1701_radio&xray}
\end{figure*}
\noindent
We place XTE J1701 on the radio -- X-ray correlation plane using quasi-simultaneous observations (within the same day) (Figure~\ref{fig:RADXray_corr}). The measurements are the pink circles representing the detection matches and we represent the radio upper limits with pink upside down triangles. A distance of 8.8\,kpc was used to determine the radio -- X-ray luminosities and assuming a flat spectral index the radio measurements were converted to 5\,GHz.  We also note that the source lies closely along the averaged Z-source points, an expected result since the source is known to transition from the Z-source to the Atoll during outbursts \citep{2009ApJ...696.1257L,2011AJ....142...34D}. 
\section{Discussion}
In this section we explore the comparisons of the 2006/7 and 2022/23 outbursts. We measure the in-band spectral energy and minimum energy from the three flares observed in the 2022/23 lightcurve. We then place the transitioning Z-Atoll source on the radio -- X-ray  correlation plane. 
\subsection{The 2006/7 outburst vs the 2022/23 outburst}
The comparison of the two known outbursts of XTE J1701 displayed in Figure~\ref{fig:J1701_radio&xray} describe the evolution of the source during those periods. T=0 corresponds to 53757.91 and 59829.65 for the 2006/7 and 2022/23 outbursts, respectively. \cite{2009ApJ...696.1257L} describes Z tracks for the 2006/7 outburst corresponding to MJD 53756.6–53767.3, 53870.5-53876.2, 53893.0–53898.0, 54112.0–54119.0, 54188.0–54196.5, 54268.5–54277.0, and 54290.0–54299.0, respectively. In the third panel of Figure~\ref{fig:J1701_radio&xray} we note that in 2006/7 the source was more active past $>$200 days after the onset of the outburst. \cite{2009ApJ...699...60L} suggested the transition to Atoll only occurred on 25 July 2007 (MJD 54306.57, more than 550 days into the outburst). As a result, we propose the source remained in the Z state throughout the 2022/23 period of our observations. In Figure~\ref{fig:J1701_radio&xray} we compare the radio lightcurves from the 2022/23 (second panel) and 2006/7 (fourth panel) outbursts. We note that more flares were observed in the 2022/23 radio lightcurve compared to the 2006/7 one, however, this is likely due to a difference in sampling since we were able to observe more consistently for a longer period. The X-ray daily averages indicate more flaring in the 2006/7 outburst (Figure~\ref{fig:J1701_radio&xray} third panel).
\begin{figure*}
    \centering
    \includegraphics[scale=0.45]{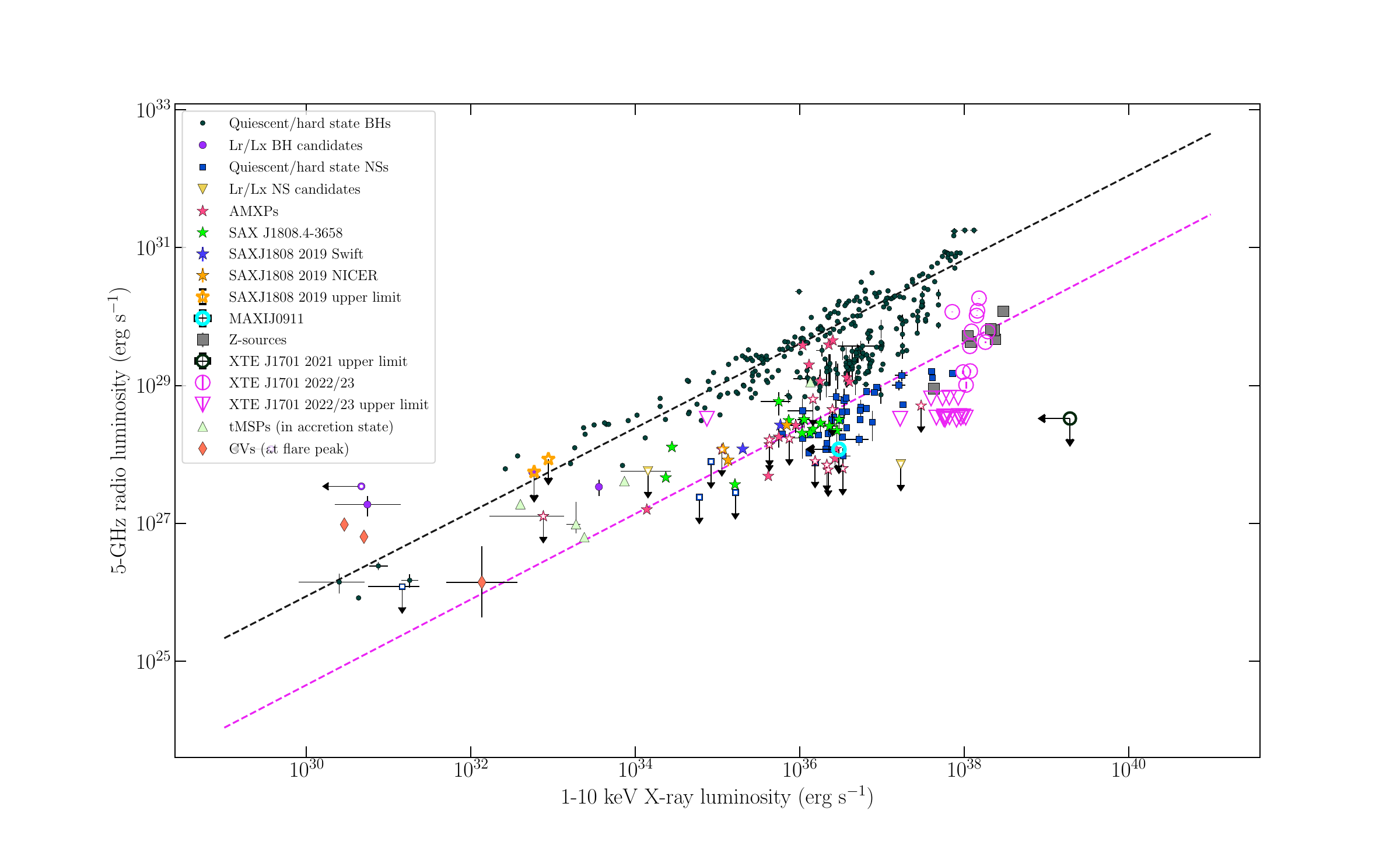}
    \caption{The radio: X-ray correlation plane, including BH and NS star sources from \protect\citealt{2006MNRAS.366...79M,arash_bahramian_2018_1252036}. The black dashed line indicates the slope for the standard track for BH systems. On the other hand, the pink slope represents the best fit (0.82 $\pm$ 2.4) for our 2022/23 XTE J1701 (pink markers) results and we find that the source lies close to the region of the six quasi-persistent Z-sources (grey squares).}
    \label{fig:RADXray_corr}
\end{figure*}
The hardness ratio was determined using energy bands $(10-20)/(2-4)$\,keV and is displayed in the third panel of Figure~\ref{fig:J1701_2022lightcurve}. We observe a pattern where the hardness ratio appears to rise with every radio flare observed. We are however, unable to 
make any strong statements on the connection between X-ray states and radio emission as our weekly radio sampling is much slower than the typical X-ray state change timescale for Z source of $\sim$ hours. 
\subsection{Polarisation constraints}
In Figure~\ref{fig:J1701_2022lightcurve} we indicate the IXPE observations Ob1 and Ob2 with grey dashed lines from \cite{2023A&A...674L..10C}. The authors observe a polarisation of 4.6 $\pm$ 0.4\,$\%$ (Ob1) on the decay of the first flare and later upper limit 1.5$\%$ (Ob2). We constrain the polarisation with 3$\sigma$ upper limits 8.6$\%$ and 8.8$\%$ for Ob1 and Ob2 respectively. During the 2006/7 outburst, \cite{2007MNRAS.380L..25F} determined upper limit constraint $<$6$\%$ for linear and circular polarisation when the source was brightest following the NB/HB track. On the other hand \cite{2023A&A...674L..10C} argue the polarisation they measure for the 2022/23 outburst is strongest in the intermediate HB track.  
In Figure 1 we indicate the IXPE observations Ob1 and Ob2 with grey dashed lines from \cite{2023A&A...674L..10C}. The authors observe a polarisation of 4.6 $\pm$ 0.4\,$\%$ (Ob1) on the decay of the first flare and later upper limit 1.5$\%$ (Ob2) and determine the X-ray polarisation they measure is strongest in the intermediate HB track. We determine the radio polarisation with 3$\sigma$ upper limits 8.6$\%$ and 8.8$\%$ for Ob1 and Ob2 respectively. In the 2006/7 outburst, \cite{2007MNRAS.380L..25F} determined an upper limit constraint $<$6$\%$ for linear and circular radio polarisation when the source followed the NB/HB track. We are however, unable to constrain the spectral state of the source for our radio polarisation upper limits during the 2022/23 outburst.
\begin{figure}
    \centering
    \includegraphics[scale=0.37]{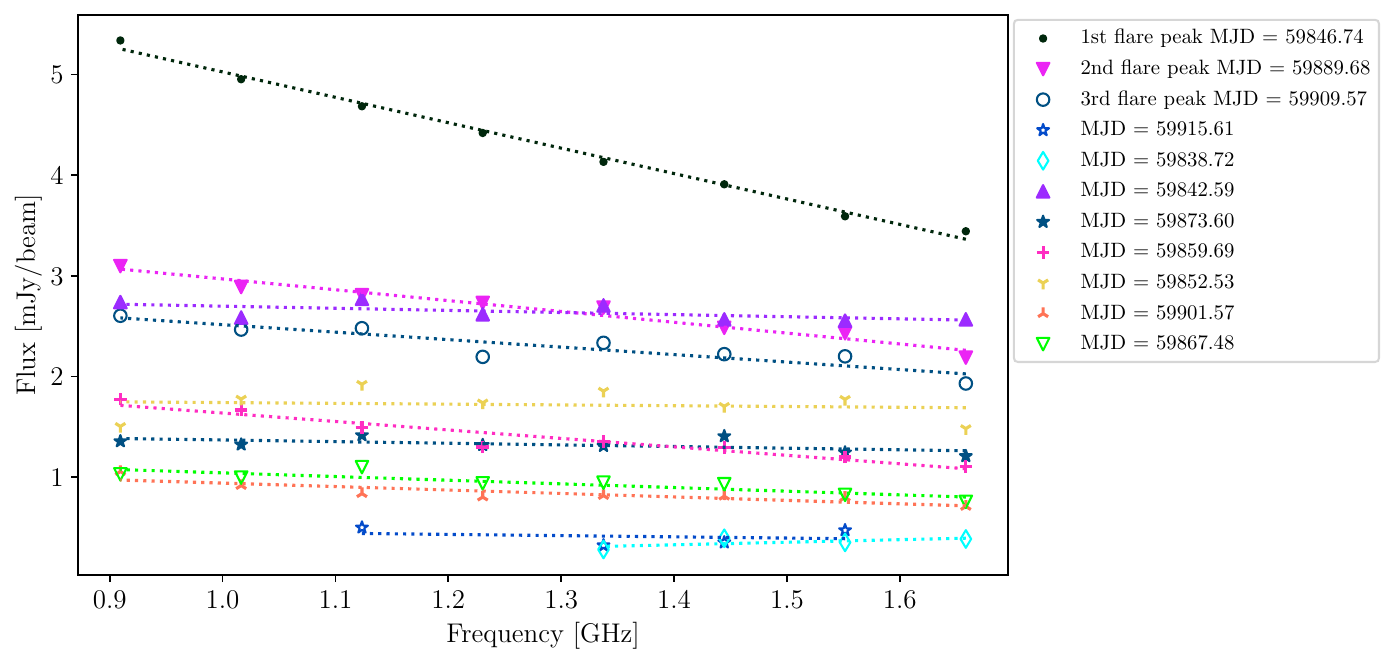}
    \caption{In-band spectra of the MeerKAT detections for the 2022/23 outburst, with slopes representing the spectral indices. The first, second and third radio flare peaks are highlighted in the legend along with the MJDs for the detections. }
    \label{fig:inbandspec}
\end{figure}
\begin{table}
\caption{The source size, minimum energy, B-field and temperature of the three flare are calculated using Equations 28, 29, 30 and 31 in \protect\cite{2019MNRAS.489.4836F}. We use the 1.28\,GHz MeerKAT frequency, the flux from each flare peak and a distance of 8.8\,kpc is assumed.}
\begin{tabular}{l c c c c}\hline
Flare & R [cm] & $E_{\rm min}$ [erg] & $B_{\rm min}$ [G] & $T_{\rm min}$ [K] \\\hline
    First & 2$\times10^{13}$ & $1\times10^{38}$ & 0.17  & $4\times10^{10}$ \\ 
   Second & 2$\times10^{13}$ & $6\times10^{37}$  & 0.18 & $4\times10^{10}$ \\
   Third & 1$\times10^{13}$ & $5\times10^{37}$  & 0.18  & $4\times10^{10}$\\\hline
\end{tabular}
\label{tab:spectab}
\end{table}

\subsection{In-band spectral index and minimum energy of the 2022/23 flares}
We measured the spectral index of the radio detections, including the peaks of the flares from the 2022/23 outburst (see Figure~\ref{fig:inbandspec}). Using scipy's linear regression to fit the best fit line, the slope for the flares are  -2.53 $\pm$ 0.20 (which we conclude is not a physical result due to the steepness of the measurement),  -1.08 $\pm$ 0.21 and -0.74 $\pm$ 0.33 for the first (MJD 59846.74), second (MJD 59889.68) and third (MJD 59909.58) flare respectively (Figure~\ref{fig:inbandspec}). 
These results suggest the flare peaks are optically thin and flatter from the first flare to the third complementing the re-ignition of the jet with the reflaring. The relatively poor sampling and hint of a flatter radio spectrum at our first detected epoch suggests we cannot rule out the flare peaks being due to the transition from optically thick to optically thin ejecta.
\\\\
Under the assumption that the peaks are due to this evolution in synchrotron self-absorption, 
the minimum energy produced at the peak of each flare can be estimated using Equation 29 of \cite{2019MNRAS.489.4836F}, the values are  $E_{\rm min} = $ $1\times10^{38}$\,erg, $E_{\rm min} = $ $6\times10^{37}$\,erg and $E_{\rm min} = $ $5\times10^{37}$\,erg for the first, second and third peak respectively. However, as noted in \cite{2023MNRAS.518.1243F} poor radio sampling can lead to a blending of multiple shorter flares and a consequent {\em underestimate} of the total output kinetic energy by over an order of magnitude. Assuming a similar scaling, which is uncertain but may be conservative given the known tendency for Z sources to produce many rapid flares, XTE J1701 is likely to have output at least a few times $10^{39}$\,erg in its jets during the outburst. Better sampling, resolution of jets to measure bulk Lorentz factor, etc, in the future are only likely to drive up such estimates. In Table~\ref{tab:spectab} we include the value of the source size R, magnetic field (B-field) and temperatures of the three flares. In a BH system like V404 Cyg the minimum energies for the flares range between $10^{38 -- 39}$\,erg and the B-field is about three times as large for V404 Cyg compared to the our results for XTE J1701 (see V404 Cyg results in \citealt{2023MNRAS.518.1243F}). \cite{2019MNRAS.489.4836F} further discuss that Cyg X-3 and GRS 1915+105 have minimum energies during flaring events that range of the order of $10^{38 -- 45}$\,erg, and the B-fields can reach as high as 2 -- 4 times our XTE J1701 results. On the other hand, the three BH systems have temperatures similar to $T_{\rm min}$ described in Table~\ref{tab:spectab} and the minimum source size of $R_{\rm{min}}$ = 7$\times10^{13}$\,cm for V404 Cyg which is about 3 times what we get for XTE J1701 ($R_{\rm{min}} = \beta_{\rm{m}}c\Delta t$ where, $\beta_{\rm{m}}$ is Equation 28 in \citealt{2019MNRAS.489.4836F} and $\Delta \rm{t}$ is the variability timescale, see also \citealt{2023MNRAS.518.1243F}). While the minimum energy for V404 Cyg is similar to XTE J1701, the rest of the BH systems indicate that a NS system like XTE J1701 has less powerful flaring than BH systems. 
\subsection{Radio -- X-ray correlation}
The radio luminosity and X-ray luminosity correlation is used as a tool to understand the connection between jets (radio) and accretion (X-rays) in X-ray binary systems (e.g. \citealt{1998A&A...337..460H,2003A&A...400.1007C,2013MNRAS.428.2500C}). While in black holes use of the plane is largely limited to sources with quasi-steady compact jets in the hard X-ray state, the sampling of neutron stars in the plane is poor enough, and their flares in general weak enough, that can be instructive to place them in the plane even if not in an equivalent of the black hole hard state (see also discussion in \citealt{2015ApJ...809...13D} and a comparison of BH and NS jet production in \citealt{2014MNRAS.443.3270M})
\\\\
The detection matches of the XTE J1701 (pink points Figure~\ref{fig:RADXray_corr}) once fit using \textsc{scipy} linear regression reveals a slope of 0.82 $\pm$ 2.4. The large error in the slope is a result of the broad scatter in the radio luminosity range. The NS systems on the plane lie nearly parallel to the BH slope and XTE J1701 follows along the NS distribution at the higher end of the luminosity range where L$_{\rm{R}}$ $>$  $10^{29}$\,$\rm erg/s$ and L$_{\rm{X}}$ $>$ $10^{38}$\,$\rm erg/s$. \cite{2013MNRAS.428.2500C} reveals that GX 339 defines the 'standard' track for BH systems on the correlation plane with slope $\approx$ 0.6 (see track indicated with the black dashed line in Figure~\ref{fig:RADXray_corr}) and BH systems that lie outside the 'standard' track are considered outliers \citep{2013MNRAS.428.2500C}. In \cite{2023MNRAS.521.2806G} we find a slope $\approx$ 0.7 for SAX J1808.4-3658 (blue and orange stars in Figure~\ref{fig:RADXray_corr}), this source lies along other NS systems (blue squares in Figure~\ref{fig:RADXray_corr}) and along the luminosity range of the BH outliers. Thus, the addition of the XTE J1701 slope may indicate BH and NS systems could follow a similar trend on the radio -- X-ray correlation plane.   
\section{Conclusions}
The second recorded outburst of XTE J1701 reveals multiple flare events in both the radio and the X-ray lightcurves. We perform a comprehensive analysis of the outburst and we compare the 2006/7 outburst to the 2022/23 outburst in both the radio and the X-ray and propose the 2022/23 outburst remained in the Z-state throughout the our observation period (lasting $\approx$ 200\,days). We determine the radio spectral index throughout the radio active period and find that the source is optically thin overall and slightly flatter from the first to the final radio flare due to the jet re-ignition during flaring. We calculate the minimum energy for the peaks of the radio flares; $E_{\rm min} = 1\times10^{38}$\,erg, $E_{\rm min} = 6\times10^{37}$\,erg and $E_{\rm min} = 5\times10^{37}$\,erg for the first, second and third peaks respectively and propose that with improved sampling the minimum energy may be $\approx$ $10^{39}$\,erg. We propose that a system like XTE J1701 has weaker flaring mechanism than BH systems because of its smaller source size, minimum energy and magnetic fields compared to BH systems during such flaring events. Furthermore, we place XTE J1701 on the radio -- X-ray correlation plane using a quasi-simultaneity of a day for the MeerKAT and MAXI observations. The source is placed in the higher end of the luminosity range, such that it occupies the same region of space as the rest of the Z-sources. We determine a slope 0.82 $\pm$ 2.4 and due to the large uncertainty we cannot define strict comparisons to the standard slope of BHs systems (slope range 0.5 -- 0.7).

\section*{Acknowledgements}
KVSG acknowledges support from the University of Cape Town and the South African NRF. 
RF would like to thank UKRI, the ERC and The Hintze Family charitable foundation for their support.
JvdE acknowledges a Warwick Astrophysics prize postdoctoral fellowship made possible thanks to a generous philanthropic donation. IMM is supported by the South African NRF and the UCT VC 2030 Future Leaders Programme.
\noindent
\\
The MeerKAT telescope is operated by the South African Radio Astronomy Observatory, which is a facility of the National Research Foundation, an agency of the Department of Science and Innovation.
We acknowledge the use of the ilifu cloud computing facility – \url{www.ilifu.ac.za}, a partnership between the University of Cape Town, the University of the Western Cape, Stellenbosch University, Sol Plaatje University and the Cape Peninsula University of Technology. The ilifu facility is supported by contributions from the Inter-University Institute for Data Intensive Astronomy (IDIA – a partnership between the University of Cape Town, the University of Pretoria and the University of the Western Cape), the Computational Biology division at UCT and the Data Intensive Research Initiative of South Africa (DIRISA).
\section*{Data Availability}
The data underlying this article will be shared on reasonable request to the corresponding author.



\bibliographystyle{mnras}
\bibliography{reference} 








\bsp	
\label{lastpage}
\end{document}